# Giant spin torque in systems with anisotropic exchange interaction

V. L. Korenev[1,2]

[1]A.F.Ioffe Physical-Technical Institute, Russian Academy of Sciences, 194021 St. Petersburg, Russia

[2]Experimentelle Physik 2, Technische Universitat Dortmund, D-44227 Dortmund, Germany

*Control of magnetic domain wall movement by the spin-polarized current looks promising for creation of a new generation of magnetic memory devices. A necessary condition for this is the domain wall shift by a low-density current. Here I show that a strongly anisotropic exchange interaction between mobile heavy holes and localized magnetic moments enormously increases the current-induced torque on the domain wall as compared to systems with isotropic exchange. This enables one to control the domain wall motion by current density $10^4 A/cm^2$ in ferromagnet/semiconductor hybrids. The experimental observation of the anisotropic torque will facilitate the integration of ferromagnetism into semiconductor electronics.*

The key condition for the integration of magnetism into electronics is the electrical control of magnetization of ferromagnetic materials. The current-induced magnetization rotation and magnetic domain wall (DW) motion are at the heart of the spin-related torque (SRT) devices, such as magnetic random access memory [1], racetrack memory [2], logic-in-memory integrated circuits [3], spin-torque oscillator [4]. According to Berger idea [5] the spin current accompanying the charge current induces SRT, which forces the domain wall movement. This idea was developed theoretically in Refs [6, 7] and a whole number of papers as is reviewed in Refs [3, 8, 9]. The SRT results from the isotropic *s-d* exchange interaction between extended charge carrier spins (*s*-electrons) and localized spins (*d*-electrons). The SRT control of DW motion by a high-density current (~$10^8$ A/cm$^2$) was realized in ambient conditions [2]. There are two main contributions into SRT. The spin transfer torque that is the in-plane torque [6] spanned by the gradient of spatially varying magnetization unit vector $\vec{m}(\vec{r})$

$$\vec{\tau}^{in} = (\vec{v}_s^e \cdot \nabla)\vec{m} \tag{1}$$

where $\vec{v}_s^e$ is the spin velocity proportional to the current density $\vec{j}$. DW moves when the current density exceeds the intrinsic threshold value, which exists even in an ideal structure, and can be very high. Another contribution represents the out-of-plane or field-like torque [7]

$$\vec{\tau}^{\perp} = \beta_e \vec{m} \times (\vec{v}_s^e \cdot \nabla)\vec{m} \tag{2}$$

where the dimensionless parameter $\beta_e = \hbar/J_e \tau_S^e$, $J_e$ is the spin splitting of electron states due to *s-d* exchange interaction and $\tau_S^e$ is the spin relaxation time of delocalized electrons. Although $\beta_e$ is usually small ($\beta_e = 0.01$ [7]), the term $\vec{\tau}^{\perp}$ is important because it controls the domain wall transport at a current density well below the intrinsic threshold. It is reminiscent of the torque due to magnetic field, hence the name "field-like torque" [3]. Ref. [8] noted that both of these torques are obtained in the adiabatic limit of electron transport, i.e. in the absence of change of linear momentum. The latter is true for classical point-like charges when the ratio of the electron de Broglie wavelength $\lambda_{Br}$ to the width $w$ of domain wall is zero. Spatially nonlocal non-adiabatic corrections $\vec{\tau}_{na}$ (small over parameter $\lambda_{Br}/w$) appear. The $\vec{\tau}_{na}$ should be taken into account on equal footing with the $\beta$-term in Eq.(2) because both of them induce the DW motion below the intrinsic threshold [8]. In any case the current flow through the DW generates the non-equilibrium spin density $\vec{\delta S}$ of s-electrons, which exerts torque on the domain wall by means of isotropic *s-d* exchange interaction. It may look surprising, but in the case of too strong exchange $\beta_e \ll 1$ the effect of current in Eq.(2) on the DW is small. In this case the non-equilibrium spin density $\vec{\delta S}$ is almost parallel to the vector $\vec{m}$, which essentially reduces the current-induced spin torque determined by the relatively small transversal $\vec{\delta S}$ components.

Here I show that a strongly anisotropic *p-d* exchange interaction of mobile heavy holes induces a field-like torque $\vec{\tau}^{an}$, which is much larger than all previously known SRTs. The anisotropy results from the strong spin-orbit interaction leading to the Ising-type exchange interaction. The key feature is the robust spin system of heavy holes, whose spin is pinned to a certain direction $\vec{e}_z$ and does not rotate in the exchange field of the ferromagnet. However, hole spins exert the spin torque (called below the anisotropic torque)

$$\vec{\tau}^{an} = \frac{1}{\beta_h}(\vec{m} \times \vec{e}_z)\left[(\vec{v}_s^h \cdot \nabla)m_z\right] \qquad (3)$$

where the parameter $\beta_h = \hbar/J\tau_S^h$, constant $J$ characterizes the heavy hole spin splitting due to *p-d* exchange interaction with magnetic atoms, and $\tau_S^h$ is the hole spin relaxation time. The anisotropic torque acts on the domain wall as a non-uniform external magnetic field parallel to the unit vector $\vec{e}_z$ and is greatly enhanced in the strong exchange limit $\beta_h \ll 1$. The characteristic current density can be decreased by 4 orders of magnitude compared to the conventional metallic systems with isotropic exchange for the same spin velocities $v_s^e = v_s^h$. This enables one to control domain wall motion by current density $10^4$A/cm$^2$ in ferromagnet/semiconductor hybrids. To detect the DW motion one could measure the local Anomalous Hall Effect (AHE), i.e. the Hall voltage across 2D hole channel of the hybrid. The DW motion through this region changes the AHE voltage, because the direction of magnetization reverses. The experimental observation of the anisotropic torque in ferromagnet-semiconductor quantum well hybrid would facilitate the integration of ferromagnetism into semiconductor (SC) electronics and the creation of all-in-one chip computer.

Consider (Fig.1a) the exchange interaction of the two-dimensional hole gas in quantum well (QW) with magnetic atoms of the nearby ferromagnetic (FM) film with perpendicular anisotropy. The QW consists of a narrow gap semiconductor layer sandwiched between two barriers of wide bandgap semiconductors. The FM is assumed to be an insulating one (e.g. ferromagnetic oxide). In this scheme the FM serves as a barrier for holes, and the wide gap semiconductor at the FM-QW interface is not necessary. The *p-d* exchange interaction of the QW holes (*p*-type system) and ferromagnet (*d*-system) results from the overlap of the hole wavefunction with FM atoms [10]. Spin–orbit interaction in combination with quantum confinement induces the splitting $\Delta_{\ell h}$ of the heavy and light hole states with total angular momentum (sum of orbital and spin momenta) projections $M_h = \pm 3/2$ and $\pm 1/2$ onto the growth direction $z$, with the heavy holes $M_h = \pm 3/2$ being the ground state. The isotropic *p-d* exchange interaction [11] couples the hole states with $M_h$ differing by unity. If the characteristic value of *J* for the exchange interaction is smaller than $\Delta_{\ell h}$, the moment projections of +3/2 and −3/2 are approximate quantum numbers. In this case, it is possible to restrict the consideration to a two-level system of +3/2 and −3/2 of the hole ground state while the exchange interaction terms changing the hole spin can be neglected. Actually it is enough to have $J \leq 2\Delta_{\ell h}$ as it comes from the direct optical measurements in II-VI Mn-based semiconductor QW CdMnTe in a saturating magnetic field [12]. The exchange interaction of heavy holes with magnetic atoms becomes strongly anisotropic provided that only the heavy-hole subband is filled [13], i.e. $\Delta_{\ell h} \geq E_F$ ($E_F$ is Fermi energy of degenerate hole gas). In this sense, the hole spin is tightly bound to the direction of the structure growth. The hole states $\pm 3/2$ can be considered as the states $\pm 1/2$ of quasiparticle with pseudospin ½ [13]. Pseudospin density of the ensemble of such quasiparticles $S_z = (n_+ - n_-)/2$, where $n_+(n_-)$ is the hole concentration with momentum projection $+3/2(-3/2)$. Total concentration of holes is $n_s = n_+ + n_-$. Below the hole spin means the hole pseudospin. In the semiclassical approximation, the exchange energy $E_{ex}$ per unit surface area at the point $\vec{r} = (x, y)$ of the plane is determined by the z-component of the mean spin density of heavy holes $\vec{S}$ and the magnetization unit vector $\vec{m}$ at the same point

$$E_{ex}(\vec{r}) = -JS_z(\vec{r})m_z(\vec{r}) \quad (4)$$

It describes the interaction of the QW heavy hole spins with the effective magnetic field parallel to z-axis and proportional to the out-of-plane magnetization component $m_z$. Larmor precession frequency of hole spins in this field

$$\vec{\Omega} = \frac{1}{\hbar}\frac{\delta E_{ex}}{\delta \vec{S}} = -\frac{Jm_z}{\hbar}\vec{e}_z \quad (5)$$

is parallel to z-axis. In equilibrium the exchange interaction induces the spin polarization of holes in the quantum well. The *p-d* exchange interaction usually (*J*>0) favors collinear orientation of the magnetization and hole spin (Fig.1a). In case of incomplete spin polarization (less than ½ per particle) the equilibrium spin density of holes

$$\vec{S}_T = -n_s\frac{\hbar\vec{\Omega}}{4E_F} = n_s s_{eq} m_z \vec{e}_z \quad (6)$$

where $s_{eq} = J/4E_F$ is the average spin of the holes in the interior of the domain. In turn, the spin-polarized holes may have significant impact on the equilibrium orientation of the magnetization of the FM. This allows the electric control of magnetic moment of the hybrid even in the absence of electric current [14].

The main goal of this paper is the current-induced torque on a DW. The current flows mainly in the quantum well because the FM is an insulating [15]. However, the hole wave function overlaps with magnetic atoms due to the tunnel effect. Therefore the current of spin-polarized holes and ferromagnetism coexist in the FM region. Anisotropic *p-d* exchange interaction of drifting holes and d-electrons will exert pressure (force) on the domain wall. If the exchange constant (taking into account tunneling) $J < 2\Delta_{\ell h}$, the angular momentum of the hole will be fixed to the axis $z$, and will not rotate in the exchange field of the ferromagnet. This greatly enhances SRT compared to the case of an isotropic s-d exchange, which preserves the projection of the electron spin along $\vec{M}$ not creating any torque. The current of heavy holes is accompanied by the spin current of z-component of the hole spin, which is equal to $\vec{J}_S = n_s \vec{V} s_0 m_z = \frac{1}{e} s_0 m_z \vec{j} d_{QW}$, where $d_{QW}$ is the QW thickness, $\vec{j}$ is the hole current density (in A/cm$^2$) related to the drift velocity $\vec{V}$ as $\vec{j} d_{QW} = e n_s \vec{V}$; $s_0 m_z$ is the spin polarization of hole current [16], $e > 0$ is a hole charge. Spin current through the domain wall creates the nonequilibrium spin density $\delta \vec{S} = \vec{S} - \vec{S}_T$ in the QW (Fig.1b). For the calculation of $\delta \vec{S}$, we note that the dynamics of holes in the semiconductor is much faster than that of the magnetization of FM. Then, for a given distribution of magnetization $\vec{m}(\vec{r})$ the stationary non-equilibrium spin density $\delta S_z$ is determined by the continuity equation

$$\frac{\partial S_z}{\partial t} = -\nabla \cdot \vec{J}_S - \frac{\delta S_z}{\tau_S^h} = -\frac{1}{e} s_0 d_{QW} (\vec{j} \cdot \nabla) m_z - \frac{\delta S_z}{\tau_S^h} = 0 \qquad (7)$$

The first term in the right-hand side of Eq. (7) describes the generation of a non-equilibrium spin due to the spin current gradient, whereas the second term is the relaxation of the system to equilibrium with a time $\tau_S^h$. Equation (7) does not contain a diffusion spin current. Thus, it is assumed that spin diffusion length is much less than its drift length. Unlike isotropic case, the precession term $\vec{\Omega} \times \delta \vec{S}$ is zero, since the generation of spin is in z-direction, so that $\delta \vec{S} \| \vec{e}_z \| \vec{\Omega}$. It follows from Eq.(7) in stationary conditions

$$\delta S_z = -s_0 d_{QW} \tau_S^h \left( \frac{\vec{j}}{e} \cdot \nabla \right) m_z, \quad \delta S_x = 0, \quad \delta S_y = 0 \qquad (8)$$

Eq. (8) shows that the nonequilibrium hole spin is tightly bound to the z-direction, and does not rotate in the exchange field of the ferromagnet.

According to Landau-Lifshitz approach [17] the hole spin system exerts the torque on a ferromagnet

$$-\gamma \frac{\delta E_{ex}}{\delta \vec{M}_S} \times \vec{m} = \gamma \frac{JS_z}{Md_{FM}} \vec{e}_z \times \vec{m} \tag{9}$$

$\gamma > 0$ is gyromagnetic ratio, $\vec{M}_S = \vec{M}d_{FM}$ is the magnetic moment per unit area, $\vec{M}$ is magnetization, $d_{FM}$ is the FM film thickness (the film is assumed to be thin $d_{FM} << w$, so that the rotation of spins throughout the thickness of the FM is coherent). To calculate the current-induced torque one should use only the non-equilibrium spin density $\delta S_z$ in Eq.(9) instead of $S_z = S_T + \delta S_z$. The equilibrium spin density $\vec{S}_T$ does not depend on the current, and should be taken into account when calculating the equilibrium distribution of the magnetization of the film FM, as discussed above. Using Eqs.(8,9) we obtain the current-induced anisotropic torque

$$\vec{\tau}^{an} = \gamma \frac{J\delta S_z}{Md_{FM}} \vec{e}_z \times \vec{m} = -\frac{J\tau_S^h}{\hbar}\left(\gamma\hbar \frac{s_0 d_{QW}(\vec{j}\cdot\nabla)m_z}{eMd_{FM}}\right)\vec{e}_z \times \vec{m} \tag{10}$$

Introducing the spin velocity $\vec{v}_s^h = \frac{\vec{j}}{e}\frac{\hbar\gamma s_0 d_{QW}}{Md_{FM}}$ and taking into account parameter $\beta_h = \frac{\hbar}{J\tau_S^h}$ one obtains the main Eq.(3). It was obtained in the semiclassical approximation, where the holes can be considered as point charges with $\lambda_{Br} = 0$. Hence the torque $\vec{\tau}^{an}$ is adiabatic. However, the direction of the hole spin is strictly bound to the axis z, and does not follow the direction of $\vec{m}$. This is the main advantage compared to the FM system with an isotropic exchange: hole spins do not turn in the exchange field of the FM, thus strongly increasing the torque for $\beta_h << 1$.

The hole spin-polarized current exerts pressure on the DW due to the *p-d* exchange. To find it let's displace the DW as a whole by a small distance and calculate the change in the *p-d* exchange energy. The DW centered at the position $x = x_0$ undergoes the pressure [18]

$$P = -\frac{\partial}{\partial x_0}\left[\int_{-\infty}^{+\infty} E_{ex}(x-x_0)dx\right] = -J\int_{-\infty}^{+\infty} S_z \frac{\partial m_z(x-x_0)}{\partial x}dx \tag{11}$$

The Eq.(11) is consistent with Ref.[19], which divides the force acting on the DW on the sum of the internal $f_{in}$ and external $f_{ex}$ forces. The force $f_{in}$ is determined by the exchange energy, the anisotropy energy, etc. The force $f_{ex}$ is determined by the external magnetic field. The latter force shifts the DW. The pressure (force according to Ref.[19]) due to anisotropic p-d exchange also contains two terms. The first term is due to the equilibrium part $S_T$ of the total spin density of holes $S_z = S_T + \delta S$. The second term originates from its non-equilibrium part $\delta S$ that exists when the current flows in the QW plane. The first term contributes to the force $f_{in}$ affecting (insignificantly in this case) the equilibrium DW structure. The second term $\sim \delta S$ defines the external force $f_{ex}$, which is responsible for the displacement of the DW as if it was an external magnetic field (Eq.(12) below). Let electric current flow in $x$-direction (Fig.1b). We assume the distribution of $\vec{m}(x)$ varies in one direction $x$ and corresponds to the Neel-type $180^0$-wall, with the z (y) axis being an easy (hard) magnetization direction. The magnetization spatial distribution is $m_z(x) = -th(x/w)$ [20]. Inserting $\delta S_z$ from Eq.(8) into Eq.(11) one obtains the current-induced pressure $P = \frac{4Js_0}{3} \frac{jd_{QW}\tau_S^h}{ew}$ (the same result is obtained for Bloch-type DW, too). The physics is straightforward. The current injects the non-equilibrium hole spin density $\delta S \sim \frac{J_S \tau_S^h}{w} \sim s_0 \frac{jd_{QW}\tau_S^h}{ew}$ into the DW region, thus increasing the exchange energy and creating the pressure $P \sim J\delta S$. It pushes the DW in such a direction as to minimize the FM-QW exchange energy: DW shifts along (opposite to) electric current for a positive (negative) product $Js_0$. The pressure can be considered [18] as a result of the action of the effective magnetic field $H_{eff}^{an}$ acting on $180^0$ DW

$$H_{eff}^{an} = \frac{P}{2M_S} = \frac{2}{3w}\left(\frac{J\tau_S^h}{\hbar}\right)\left(\frac{jd_{QW}s_0\hbar}{eMd_{FM}}\right) = \frac{1}{\gamma\beta_h}\frac{2v_s^h}{3w} \quad (12)$$

The same result one obtains by substituting the anisotropic torque Eq. (3) into the Landau-Lifshitz-Gilbert equation for DW motion below the Walker breakdown in a way similar to Ref.[8] for isotropic case. Thus, the anisotropic torque Eq.(3) is the field-like torque, which works similar to the beta-term Eq.(2). It should be noted that, the pressure $P$ is not literally the pressure $P_H = 2M_S H$ produced on DW by the external magnetic field $H$. The transformation $M_S \to -M_S$ (which transforms DW into successive DW) changes the sign of $P_H$ but not the sign of $P$. In the former case the DW reverses the direction of motion and doesn't reverse it in the latter case. The successive DWs will move in the same direction, which can be changed by the electric current direction. Correspondingly, unlike the field $H$, the effective field Eq.(12) changes its sign under magnetization inversion.

It is instructive to compare the field $H_{eff}^{an}$ with the effective field $H_{eff}^{iso}$ produced by the field-like torque $\vec{\tau}^\perp$ in the metallic ferromagnets with isotropic exchange. The result is $H_{eff}^{iso} = \frac{\beta_e v_s^e}{\gamma w}$ [7], which is $1/\beta_e \beta_h$ times smaller provided that the spin velocities are the same, $v_s^e = v_s^h$. It is shown below that the gain $1/(\beta_e \beta_h)$ can be $10^4$. Thus the field $H_{eff}^{an}$ by 4 orders of magnitude larger than the field $H_{eff}^{iso}$. The reason is that the non-equilibrium spin density $\vec{\delta S}$ in isotropic case is almost parallel to the vector $\vec{m}$ due to the fast rotation in the exchange field of the ferromagnet, which essentially reduces the current-induced spin torque $\vec{\delta S} \times \vec{m} \to 0$. In contrast to it, the strong exchange anisotropy bounds tightly the non-equilibrium spin $\vec{\delta S}$ to the z-direction, thus greatly increasing the torque. As a result the typical current density is reduced from $10^8$ A/cm$^2$ [2] in metals down to $10^4$ A/cm$^2$ and less in the semiconductor-based heterostructures. In case of intermediate anisotropy one should add into Eq. (4) the term $-J_\perp (S_x m_x + S_y m_y)$. Torque enhancement will take place if the exchange anisotropy is not too small, i.e. $|J - J_\perp| \sim J$. Detailed analysis of the general case is out of scope of this paper.

One has to fulfill the following conditions for realization of a giant anisotropic torque: (1) a strong spin-orbit interaction in the valence band $\Delta_{so} > J$; (2) its anisotropy, fixing the angular momentum of hole along a preferred axis, $\Delta_{\ell h} > J/2, E_F$ ; (3) long spin relaxation time $\tau_s^h \gg \hbar/J$ to get $1/\beta_h \gg 1$.

All of these conditions can be implemented in semiconductor QWs near the FM thin film. For example, $\Delta_{so} \approx 0.3\,eV$ and $\Delta_{\ell h} \leq 0.05\,eV$ are in GaAs-type semiconductor quantum wells. The concentration $n_s$ of holes filling the heavy-hole subband follows from the inequality $\Delta_{\ell h} > E_F$ [21], where the Fermi energy of two-dimensional hole gas (2DHG) can be estimated as $\frac{\pi \hbar^2}{m_h} n_s$. Using the in-plane hole mass $m_h = 0.2 m_0$ ($m_0$ is the free electron mass) one gets $n_s < 4 \times 10^{12}\,cm^{-2}$. The spin relaxation time of QW holes can be suppressed up to 10 ps at room temperature [22]. This gives the energy window for $J$: $2\Delta_{\ell h} \approx 0.1\,eV > J > \hbar/\tau_S^h \approx 5 \times 10^{-5}\,eV$. The $J$ value can be varied in a wide range by the FM layer thickness. It can be estimated as $J \approx J_{FM} D$, where $J_{FM}$ is the exchange splitting inside of FM (up to a few eV), $D$ is the transmission coefficient of the FM layer. For D=0.002 (a few nm thick FM) one has $J \approx 5\,meV$. Hence $1/\beta_h = 100$, thus giving the gain $1/(\beta_e \beta_h) = 10^4$. The spin velocity $v_s^h$ in FM-QW hybrid is comparable to that $v_s^e$ for metallic systems. Indeed, $v_s^e = 10^4$ cm/s for the current density $j_e = 10^8$ A/cm$^2$ [8]. Estimation according Eq.(10) for $d_{QW} = d_{FM}$, $M/\gamma\hbar = 10^{22}$ cm$^{-3}$, $s_0 = 0.1$ gives $v_s^e = v_s^h$ for nominally the same current density. Thus the QW hole current $j$, giving the same torque value as in metallic FMs, is reduced down to $j \approx \beta_e \beta_h j_e = 10^4$ A/cm$^2$ in the FM-QW hybrid. DW motion can be detected in a conventional technique (Fig.2). One could measure the local Anomalous Hall Effect (AHE), i.e. the Hall voltage $V_{AHE} \sim j m_z$ across 2D hole channel. The DW motion through the local region changes the voltage, because the direction of $\vec{m}$ reverses.

Deriving Eq.(12) we assumed the Neel (Bloch) DW profile being independent from the FM-SC coupling strength. Generally the FM and SC may form the strongly coupled spin system as it was argued previously in [10, 14]. However in the case under study the exchange energy $E_{FM}^{ex}$ and anisotropy energy $E_{FM}^{an}$ (both per unit surface) of the FM itself are much stronger than the FM-SC anisotropic exchange coupling energy $E_{ex}$, Eq.(4). Thus the latter can be considered as a small perturbation not affecting the domain wall structure. Indeed, for typical values of Curie temperature $T_c \sim 0.1\ eV$ (in energy units), the concentration of magnetic atoms $N \sim 10^{22}$ cm$^{-3}$, the FM film thickness $d_{FM} = 5\ nm$ and the anisotropy constant $K = 10^5 - 10^6\ erg/cm^3$ [20], one obtains $E_{FM}^{ex} \cong T_c N d_{FM} = 5 \times 10^{14}\ eV/cm^2$, $E_{FM}^{an} \cong K d_{FM} = 3 \times 10^{11}$-$3 \times 10^{12}\ eV/cm^2$. These energies are much larger than the FM-SC interface coupling energy $E_{ex} \cong J n_s = 5 \times 10^9\ eV/cm^2$ for $J = 5\ meV$ and hole concentration in the QW $n_s = 10^{12}\ cm^{-2}$. This justifies the assumption of independency. At the same time the exchange coupling constant as low as $J = 5\ meV$ is enough to induce giant torque enhancement, because the most important parameter of the problem is $1/\beta_h = J\tau_S^h/\hbar \gg 1$.

Anisotropic torque Eq.(3) and the effective field Eq.(12) associated with it can be realized in a media of rather high symmetry - the cylindrical symmetry with inversion center. New torque – the spin-orbit torque - arises in systems without inversion center, where there is the linear-in-momentum spin-orbit splitting of bands. The current-induced spin-orbit torque on spin-polarized electrons was discovered in non-magnetic quantum well [23]. In this case, the spin current directly induces a non-equilibrium spin density of QW carriers $\delta S' \propto J_S \tau_S / \ell_{so}$, where characteristic length $\ell_{so} = \hbar/mQ$, $m$ – effective mass; parameter $Q$, having velocity dimension, is a measure of the spin-orbit coupling strength [24]. The spin density $\delta S'$ will be comparable to $\delta S \propto J_S \tau_S / w$ from Eq.(7), when the length $\ell_{so} \leq w$. Recently Ref. [25] demonstrated that the *p-d* exchange interaction of holes with magnetic atoms together with spin-orbit coupling in the valence band creates the spin-orbit torque on DW in magnetic semiconductor GaMnAs. Also it can be expected that the anisotropic *p-d* exchange of FM with QW holes will produce an additional pressure $P' \sim J \cdot \delta S'$ on DW due to the spin-orbit torque.

The idea of spin torque enhancement goes beyond FM-SC hybrids. The anisotropy of *p-d* exchange [26] can occur in strained ferromagnetic GaMnAs, InMnAs. In contrast to FM-SC hybrids here the anisotropic *p-d* exchange determines both the domain wall structure and the current-induced effective magnetic field for GaMnAs (InMnAs) samples. However, the Eq. (11) is still applicable because it is quite general. The Eq.(11) was written in Ref.[18] for isotropic exchange in a way similar to the calculation of force in open systems [27]. Later Ref.[6] came to a similar result (Eq.6 in [6]) from microscopic approach. Ref.[8] also noted the general character of the equation like Eq.(11) (Eq.(136), page 240). In the case of strained magnetic semiconductor the calculation using Eq.(11) may give an additional numerical coefficient in Eq.(12) of the order of one. Therefore the large effective magnetic field $H_{eff}^{an}$ can be achieved in a strained p-type ferromagnetic semiconductor GaMnAs (InMnAs). It was noted earlier that the strong spin-orbit interaction affects the SRT [28, 29, 30]. It increases the torque $\bar{\tau}^{\perp}$ partly due to the fact that the parameter $\beta_h$ is close to 1. However, the role of the anisotropy of the *p-d* exchange, which, as we have seen, changes the concept, was not discussed. A thin strained GaMnAs film with a significant splitting of the light and heavy hole states is necessary to obtain the strong *p-d* exchange anisotropy. The splitting can be estimated [31] as $\Delta_{\ell h} = 6.2\varepsilon = 0.1\ eV$ for a strain $\varepsilon = 1.5\%$ [32]. The spin relaxation time of holes (*0.1 ps* in cubic GaAs) becomes longer in strained semiconductors [33]. Using as a reasonable estimation $\tau_S \approx 1\ ps$, one has $J > 0.6\ meV$. The upper bound of *J* depends on Mn concentration. Theoretical estimations give the heavy-hole spin splitting 0.2 eV [34] – 0.5 eV [31] for Mn concentration $x \approx 5\%$. Hence $J \leq 2\Delta_{\ell h} = 0.2$ eV for $x \approx 2-3\%$. For *J=0.1 eV* the exchange anisotropy is large and the parameter $1/\beta_h > 100$, so the gain is $1/(\beta_e \beta_h) > 10^4$. Additionally, the unique property of GaMnAs is a very small magnetization value (~1/10 of *M* of the ordinary FMs), which increases the spin velocity [8]. As a result, the driving current value further decreases down to $10^3$ A/cm$^2$ in strained GaMnAs in contrast to $10^8$ A/cm$^2$ in metallic FMs.

Thus the anisotropic *p-d* exchange interaction of mobile heavy holes and localized magnetic moments creates an anisotropic torque on the domain wall, which is much larger than all known spin related torques in a ferromagnet with isotropic *s-d* exchange. This enables the control of the domain wall movement by the current density as low as $10^4$ A/cm$^2$. The current of such a density can flow even in a semiconductor, without damaging it. This can shift the magnetic memory paradigm from the purely metallic systems to a wide class of the ferromagnet-semiconductor hybrids. The experimental observation of the anisotropic torque will facilitate the integration of ferromagnetism into semiconductor electronics and creation of the all-in-one chip computer.

This work was supported by the Deutsche Forschungsgemeinschaft within the Gerhard Mercator professorship program, the RFBR, Program of RAS, and the Government of Russia (project 14.z50.31.0021, leading scientist M.Bayer).

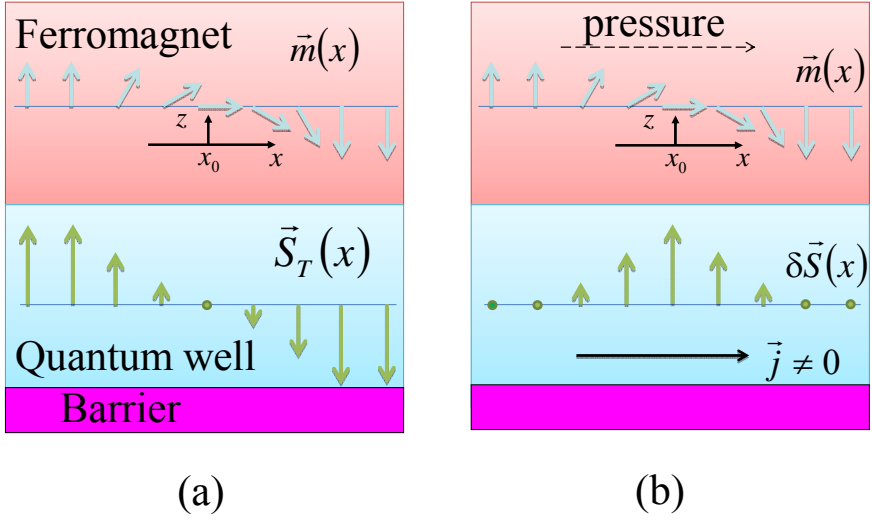

Figure 1 (color online). Ferromagnet (FM) – Semiconductor quantum well (QW) heterostructure. (a) Electric current is absent. Distribution of magnetization $\vec{m}(x)$ of the FM (blue arrows in the up rectangle) within the one-dimensional $180^0$-Neel domain wall. Equilibrium distribution $\vec{S}_T(x)$ of the heavy hole spin density (green arrows in the bottom rectangle). (b) Current, $\vec{j} \parallel x$, flows. The distribution of non-equilibrium spin density $\delta\vec{S}(x)$ of the heavy holes (green arrows in the bottom rectangle) injected by spin-polarized current. The domain wall undergoes pressure due to *p-d* exchange interaction between FM-QW spin systems, thus forcing DW to the right (dashed line in the up rectangle) for a positive $Js_0$. Easy (hard) magnetization direction coincides with the z(y) axis.

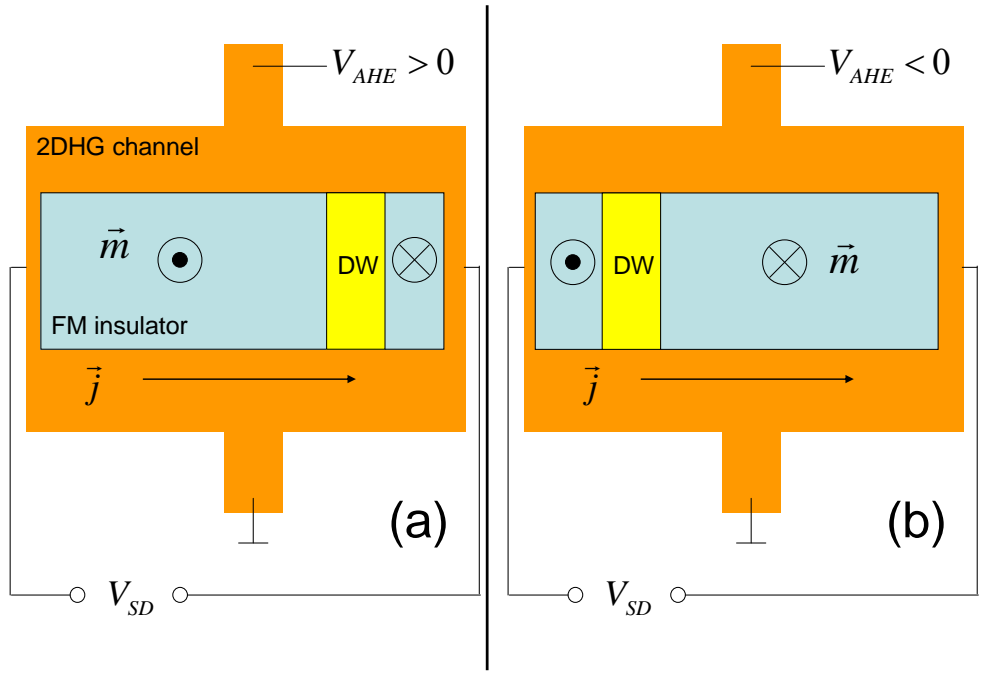

Figure 2 (color online). The magnetization readout is made by the local measurement of the AHE voltage $V_{AHE} \sim jm_z$ in the small current regime $j < j_{threshold}$, when the DW is static. The DW position is controlled by the pulses of the source-drain voltage $V_{SD}$. The DW motion through the Hall bars region reverses the sign of $V_{AHE}$. Voltage ($V_{SD}$) pulse moves the DW (a) to the right with $V_{AHE} > 0$ corresponding to bit 1; (b) to the left with $V_{AHE} < 0$ corresponding to bit 0.